\documentclass[acmsmall, 11pt, nonacm=true]{acmart}
\usepackage{breakurl}
\usepackage{tabularx}

\AtBeginDocument{%
  \providecommand\BibTeX{{%
    \normalfont B\kern-0.5em{\scshape i\kern-0.25em b}\kern-0.8em\TeX}}}
    
\copyrightyear{2022}
\acmYear{2022}
\setcopyright{rightsretained}
\acmConference[FAccT '22]{2022 ACM Conference on Fairness, Accountability, and Transparency}{June 21--24, 2022}{Seoul, Republic of Korea}
\acmBooktitle{2022 ACM Conference on Fairness, Accountability, and Transparency (FAccT '22), June 21--24, 2022, Seoul, Republic of Korea}\acmDOI{10.1145/3531146.3533228}
\acmISBN{978-1-4503-9352-2/22/06}

\acmSubmissionID{917}

\begin{document}

\title{Adversarial Scrutiny of Evidentiary Statistical Software}

\author{Rediet Abebe}
\affiliation{%
    \institution{University of California, Berkeley}
    \country{USA}
}

\author{Moritz Hardt}
\affiliation{%
    \institution{Max Planck Institute for Intelligent Systems, T\"ubingen} 
    \country{Germany}
}

\author{Angela Jin}
\affiliation{%
    \institution{University of California, Berkeley}
    \country{USA}
}

\author{John Miller}
\affiliation{%
    \institution{University of California, Berkeley}
    \country{USA}
}

\author{Ludwig Schmidt}
\affiliation{%
    \institution{University of Washington}
    \country{USA}
}

\author{Rebecca Wexler}
\affiliation{%
    \institution{University of California, Berkeley}
    \country{USA}
}

\begin{abstract}

The U.S. criminal legal system increasingly relies on software output to convict and incarcerate people. In a large number of cases each year, the government makes these consequential decisions based on evidence from statistical software---such as probabilistic genotyping, environmental audio detection, and toolmark analysis tools---that defense counsel cannot fully cross-examine or scrutinize. This undermines the commitments of the adversarial criminal legal system, which relies on the defense’s ability to probe and test the prosecution’s case to safeguard individual rights.

Responding to this need to adversarially scrutinize output from such software, we propose \emph{robust adversarial testing} as an audit framework to examine the validity of evidentiary statistical software. We define and operationalize this notion of robust adversarial testing for defense use by drawing on a large body of recent work in robust machine learning and algorithmic fairness. We demonstrate how this framework both standardizes the process for scrutinizing such tools and empowers defense lawyers to examine their validity for instances most relevant to the case at hand. We further discuss existing structural and institutional challenges within the U.S. criminal legal system that may create barriers for implementing this and other such audit frameworks and close with a discussion on policy changes that could help address these concerns.

\end{abstract}

\begin{CCSXML}
<ccs2012>
   <concept>
       <concept_id>10010405.10010455.10010458</concept_id>
       <concept_desc>Applied computing~Law</concept_desc>
       <concept_significance>500</concept_significance>
       </concept>
   <concept>
       <concept_id>10010147.10010257</concept_id>
       <concept_desc>Computing methodologies~Machine learning</concept_desc>
       <concept_significance>300</concept_significance>
       </concept>
   <concept>
       <concept_id>10011007</concept_id>
       <concept_desc>Software and its engineering</concept_desc>
       <concept_significance>300</concept_significance>
       </concept>
 </ccs2012>
\end{CCSXML}

\ccsdesc[500]{Applied computing~Law}
\ccsdesc[300]{Computing methodologies~Machine learning}
\ccsdesc[300]{Software and its engineering}

\keywords{adversarial scrutiny, algorithmic accountability, algorithmic audit, black-box testing, evidentiary software, statistical software, robust machine learning}

\maketitle

\section{Introduction}

The U.S. criminal legal system increasingly uses output from software tools to prosecute the criminally accused. Statistical software is especially gaining prominence and has been deployed in a range of settings, from probabilistic genotyping to environmental audio detection and forensic bullet analysis. At present, while the government uses output from such software as evidence, defense counsel are often unable to fully cross-examine or scrutinize these tools. This limitation, coupled with the accelerated use of machine learning and statistically-driven software, presents a clear and growing danger to the legal truth-seeking process.

The U.S. uses an adversarial system, where law enforcement is tasked with investigating evidence of guilt, and defense counsel is tasked with probing the prosecution’s case and investigating evidence of innocence. By pitting these two roles against one another, the system as a whole seeks legal truth and the protection of individual rights. In the context of evidentiary software, however, numerous factors prevent defense counsel from effectively performing their duty to scrutinize the prosecution’s case.

One such factor is uncertainty over what adversarial scrutiny of a software program entails~\cite{engstrom2020legal}. The U.S. Constitution guarantees criminal defendants the right to confront and cross-examine the witnesses against them~\cite{amend6}. Yet, courts, counsel, and legal scholars have fractured over how and whether to apply these rights to output from software~\cite{2012people}. Compounding this uncertainty, there are no technical standards to adversarially scrutinize output from software used as evidence of guilt. Existing mechanisms for doing so are insufficiently powerful and incomprehensive to safeguard against erroneous output. 

In this work, we tackle this issue of evaluating the validity of evidentiary statistical software by proposing a technical conceptual audit framework—\emph{robust adversarial testing}. We draw on a growing body of work in robust machine learning to formally define robust adversarial testing and describe how to operationalize it within the criminal legal system. This framework connects the notion of adversarialism in the U.S. legal system with adversarial robustness in machine learning. The connection is of mutual benefit: Defense counsel can repurpose tools and insights from robust machine learning in service of adversarial scrutiny. The specific requirements of defense counsel, in turn, contribute well-motivated research and engineering opportunities to the statistics and machine learning communities. In doing so, this framework provides a process that is concrete enough to be adopted for adversarial scrutiny of statistical software across the legal system and flexible enough to empower the defense to adversarially select inputs for testing on a case-by-case basis.

Specifically, our proposal of robust adversarial testing rests on two components: First, the defense can adversarially select a family of inputs to the statistical software that is most relevant to the defendant’s case. This could entail, for instance, testing the software’s performance for specific demographic groups. It also allows the defense to adversarially select relevant performance measures, such as the rate of false inclusion of the defendant in a forensic sample. Through these two levers, the defense may adversarially test the validity of the statistical software on the inputs most relevant to the case at hand.

In addition to providing a framework for adversarially scrutinizing evidentiary statistical software, our proposal adds to and complements a growing body of work in the accountability community that tackles challenging tasks around auditing algorithmic and software tools used at various stages of the U.S. criminal legal system. While our motivation in this work is evidentiary software, the framework here can apply more broadly to statistical software used at any stage of a legal case. Likewise, our work and discussions presented here focus on the U.S. criminal legal system, but the framework may prove useful in other geographic settings. As such, this work takes a step towards systematizing and formalizing a key component of audits of statistical software used in the law more generally.

Our work also goes beyond existing research in algorithmic audits and accountability in its focus on evidentiary software. Existing work thus far has primarily focused on tools predominantly used at the investigation, pretrial, and sentencing stages, such as risk assessment tools and facial recognition technology~\cite{angwin2016machine, buolamwini2018gender, chouldechova2017fair, corbett2018measure, dressel2018accuracy, eckhouse2019layers, huq2018racial, kehl2017algorithms, kleinberg2016inherent, koepke2018danger, mayson2018bias, osoba2017intelligence, raji2019actionable, slobogin2021just}. While output from these tools may at times be used as evidence—and indeed there is a credible concern that they will be widely adopted as evidentiary tools—this is not their current primary use. As a result, the algorithmic accountability literature has not kept up pace with evaluating evidentiary software. In presenting our adversarial scrutiny framework, we highlight a path forward for the machine learning and statistics research communities to consider evaluating a broader range of software tools and to contribute resources that could empower defense counsel to more thoroughly scrutinize evidence of guilt. 

The rest of this paper is structured as follows: In Section~\ref{sec:legal-background}, we set the stage for our framework by introducing the role of adversarial scrutiny in the U.S. criminal legal system. In Section~\ref{sec:prob-genotyping}, we provide a brief introduction to probabilistic genotyping software, which we use as a running example throughout the remainder of the paper. In Section~\ref{sec:limitations}, we consider existing mechanisms for scrutinizing evidentiary software and lay out their limitations in assessing statistical software in particular. We present our proposal in Section~\ref{sec:framework}. We define and introduce adversarial scrutiny of evidentiary statistical software and its operationalization in the U.S. criminal legal system. While our primary contribution is to conceptualize the legal concept of adversarial scrutiny in a technical formulation, we acknowledge that existing structural and institutional barriers may prevent defense counsel from implementing this and other such audit frameworks in practice. We describe some of these challenges that the law and policy communities should simultaneously tackle in Section~\ref{sec:barriers}.

\section{Legal background: adversarial scrutiny in the law}
\label{sec:legal-background}
In this section, we introduce the reader to the concept of adversarial scrutiny in the law. We begin with a general commentary about adversarialism and explain why criminal defense adversarial scrutiny is essential for forensic evidence in particular. We then describe some of the legal rules and procedures that are designed to facilitate this scrutiny at three different stages of a criminal case: \emph{plea negotiations, admissibility hearings,} and \emph{trial}. Importantly, these examples are meant to be illustrative and not comprehensive. Criminal defense counsel should perform adversarial scrutiny at \emph{all} stages of a case.

The U.S. uses an adversarial criminal legal system. This means law enforcement, including prosecutors and the police, are responsible for investigating evidence of guilt. Criminal defense counsel, on the other hand, are responsible for investigating evidence of innocence and, crucially, for identifying flaws and weaknesses in the prosecution’s evidence. Criminal defense counsel are the sole actors in the entire legal system with a duty to probe, test, and challenge the prosecution’s evidence~\cite{wexler2021privacy}. We call this duty \emph{adversarial scrutiny}.

Defense counsel are supposed to perform adversarial scrutiny not as neutral inquisitors, but rather as “zealous advocates” for the criminally accused~\cite{smith1999defending}. In this sense, defense counsel are similar to white hat hackers who are supposed to focus on finding flaws and weaknesses in a computer system. At trial, both the prosecution and defense present their evidence and arguments to the jury, which ultimately decides whether the prosecution has proven the defendant’s guilt beyond a reasonable doubt. Similarly, in pretrial and post-conviction proceedings, both sides present evidence and arguments to the judge, who ultimately decides the legal issues in the case. By pitting the prosecution and defense advocates against one another, the system as a whole seeks legal truth and the protection of individual rights~\cite{ogletree1992beyond}.

Adversarial scrutiny of forensic evidence can be at once technically complex and exceptionally urgent. Invalid forensic evidence has contributed to a substantial portion of the wrongful convictions of individuals who were later exonerated~\cite{garrett2009invalid, innocence_project}. Indeed, prior studies have identified invalid forensics as the "second most common contributing factor" in wrongful convictions~\cite{west2015innocence}.  It is therefore essential for defense counsel to perform effective adversarial scrutiny when prosecutors use evidentiary statistical software as forensic evidence.

The need for criminal defense counsel to perform adversarial scrutiny of forensic evidence can apply at any stage of a case. For example, the Sixth Amendment constitutional right to effective assistance of counsel during pretrial \emph{plea negotiations} may require defense counsel to independently investigate and test the prosecution's forensic evidence early in a case~\cite{garrett2016constitutional}.\footnote{We thank Brandon Garrett for highlighting the significance of the Sixth Amendment right to effective assistance of counsel in establishing a constitutional basis for defense testing of prosecution forensic evidence, especially pre-plea.} Accordingly, as soon as possible after a case begins, the defense should demand access to relevant information about any evidentiary statistical software via statutory discovery and constitutional due process \emph{Brady} rules, which require the prosecution to disclose certain information to the defense~\cite{giannelli1991criminal}, as well as by serving pretrial subpoenas on developers or other third parties for additional relevant information that the prosecution may not know or possess~\cite{giannelli1991criminal}. These discovery and subpoena rules should enable the defense to obtain information to perform adversarial scrutiny at this and subsequent stages of the case.      

As another example, the need for adversarial scrutiny may also arise during pretrial \emph{admissibility hearings}, in which the prosecution and defense argue before a judge about what evidence they should be allowed to present to the jury at an upcoming trial. Based on the arguments from both sides, the judge will determine whether the evidence satisfies the requirements for admissibility, which include relevance~\cite{evidence402} and lack of unfair prejudice~\cite{evidence403}, among other considerations.\footnote{See Fed. R. Evid. 401 \& 403.} For expert testimony based on evidentiary statistical software, the rules of evidence require an additional showing that the software is a method that has been reliably applied in the defendant's case~\cite{evidence702}. 

There are two leading standards across federal and state jurisdictions for this additional showing: \emph{Daubert} and \emph{Frye}~\cite{cheng2005does}. \emph{Daubert} gives judges flexibility in how they choose to make the reliability determination, but encourages them to consider the following series of factors: ``whether the theory or technique in question can be (and has been) tested, whether it has been subjected to peer review and publication, its known or potential error rate and the existence and maintenance of standards controlling its operation and whether it has attracted widespread acceptance within a relevant scientific community''~\cite{1993daubert}. The test in \emph{Frye} jurisdictions focuses on one factor: whether a method is ``sufficiently established to have gained general acceptance in the particular field in which it belongs''~\cite{1963frye}. Critics have castigated judges for failing to perform their proper gate-keeping function to keep unreliable forensics out of court, particularly in criminal cases~\cite{sinha2021radically, garrett2017myth, hilbert2018disappointing, groscup2002effects, gertner2010commentary, kaye2017daubert}. Such failures emphasize the urgent need for effective defense adversarial scrutiny in admissibility hearings so as to better educate judges about the flaws in prosecution forensic evidence and persuade them to do a better job of excluding faulty forensics from trial.  

If the defense loses an admissibility challenge and the judge permits the prosecution to present forensic evidence to the jury, then this brings us to our third example of a point in time for adversarial scrutiny: \emph{trial}. At trial, defense counsel should tell the jury about any flaws and weaknesses in the prosecution's evidence. Even if the \emph{Daubert} and \emph{Frye} standards were applied correctly, much imperfect evidence would still be admitted at trial because the requirements for admissibility in the rules of evidence are relatively lenient, including for expert testimony~\cite{bernstein2015defending}. This liberal thrust of the rules in favor of admissibility reflects an overarching policy preference for trusting juries to properly weigh evidence, including by discounting less-reliable evidence, rather than empowering judges to keep that evidence entirely secret from the jury. Further, trial courts have substantial discretion on admissibility determinations and any potential errors in their judgments are reviewed under a deferential standard by the appellate courts~\cite{saltzburg2006federal}. As a result, trials can be flooded with under-scrutinized forensic evidence that defense counsel should contest before the jury.

Several legal rules and procedures facilitate such contestations at trial. For instance, the Sixth Amendment to the U.S. Constitution guarantees criminal defendants the right to confront and cross-examine the witnesses against them~\cite{amend6}, including prosecution witnesses who testify during trial based on the output of evidentiary software~\cite{2009melendez, 2011bullcoming, 2012williams}. Legal scholars and courts are currently struggling to determine whether, to what extent, and how this right may also entitle defendants to directly cross-examine and confront evidentiary software systems themselves~\cite{cheng2018beyond, roth2016machine}. Some have suggested that lawmakers or the Constitution might require software developers, designers, or operators to take the witness stand~\cite{2021state, roth2016machine}. Others have advocated for defense access to source code, or defense access to “enhanced discovery” that would afford increased opportunities to test prosecution software~\cite{cheng2018beyond}. To date, however, few have offered technical details about what the defense should do with this information after they obtain it.

In this paper, we present a technical audit framework to operationalize defense adversarial scrutiny of evidentiary statistical software. As the preceding paragraphs explain, the criminal legal system depends on an adversarial framework that tasks defense counsel with contesting the prosecution's forensic evidence. The existing legal adversarial framework motivates our technical proposal, which we present in detail in Section~\ref{sec:framework}. We argue that defense adversarial scrutiny of evidentiary statistical software, or careful examination directed to identifying flaws or weaknesses in the software output, is crucial to protect defendants' rights and to the truth-seeking goals of the criminal legal system as a whole.

\section{A Case Study: Probabilistic Genotyping Software}
\label{sec:prob-genotyping}

In this section, we discuss probabilistic genotyping software (PGS) tools, a popular class of evidentiary statistical software, which we will use as a running example to ground the discussions and definitions that follow. 

PGS tools are widely used in criminal cases. According to the creator of STRmix, this tool has been used in over 220,000 cases worldwide~\cite{strmixUse}. Likewise, the Forensic Statistical Tool (FST), was used in over 1,300 cases from 2011 to 2017~\cite{fstUse}. Like many statistical decision-making tools, at their core, PGS tools implement a likelihood ratio test~\cite{coble2019probabilistic, hardtrecht}. As such, though we focus on PGS for concreteness, our discussion and approach apply to a much broader class of evidentiary software based on statistical techniques. Note, in discussing these tools, our goal here is not to take a stance on the validity of these tools in any specific instance but rather to provide a concrete framework to scrutinize these tools.

Roughly, forensic DNA analysis begins with a biological sample of interest obtained from a crime scene and a known reference sample taken from the defendant. The observed crime scene sample varies in complexity from simple settings involving DNA from a single person to complex cases involving a mixture of DNA from multiple contributors, e.g., a sample from a door handle that was potentially touched by multiple people. We commonly refer to the observed sample as a ``mixture.'' From these samples, an analyst extracts DNA profiles using standard laboratory techniques. PGS tools take, as input, both the observed DNA profile from the crime scene and the reference DNA profile from the defendant. They then attempt to estimate a likelihood ratio that compares the probability of observing the crime scene DNA if the reference candidate were a contributor to the mixture versus a random (unrelated) individual. PGS tools rely on, among other things, population genetic databases and laboratory-specific parameters (e.g., probability of allele drop-in) to estimate this likelihood ratio. Specific details are beyond the scope of this paper, and we refer the interested reader to~\citet{coble2019probabilistic, butler2021dna, matthews2020trusted} and the references therein for more background.

Despite their widespread use, the validity of PGS tools is an ongoing research direction and the source of much debate~\cite{lander2016forensic, butler2021dna, perlin2018dna, matthews2020trusted, matthews2019right}. For instance, in a high-profile 2016 case, two different leading PGS tools produced different conclusions when used to analyze the same DNA sample ~\cite{lander2016forensic,president2017addendum}. Many of the questions around validity stem from the fact that it depends on the complexity of the samples and the specific settings in which it is used~\cite{lander2016forensic, butler2021dna,matthews2020trusted}. Concretely,~\citet{butler2021dna} detail more than 20 distinct factors which all influence the performance and validity of PGS tools. These factors include, among others, (i) the number of contributors to a mixture, (ii) the amount of DNA collected, (iii) the relative proportion of material from each contributor, (iv) the quality of the genetic material, (v) the presence of close relatives in the sample, and (vi) calibration of software parameters to laboratory equipment. In another study,~\citet{matthews2020trusted} show how variations in the parameters of the underlying statistical model, e.g., a parameter summarizing the genetic diversity in the population, can cause substantial changes in the likelihood ratios produced by different tools. Understanding the tool's validity in any particular setting, therefore, requires understanding how these factors influence its performance.

\section{Limits of Current Assessment Mechanisms}
\label{sec:limitations}
At present, courts, counsel, and commentators have discussed a variety of mechanisms for assessing the validity of evidentiary software that defense counsel may apply to statistical tools. We focus our discussion on three frequently considered mechanisms: \emph{source code review, validation studies} by relevant scientific communities, and \emph{direct testing} by the defense. In this section, we consider how defense counsel might face limitations when using each mechanism to scrutinize the validity of statistical software.\footnote{Given our specific focus on statistical software, we do not aim to present a comprehensive list of limitations for each mechanism.} We focus our discussion on PGS tools to provide specific examples, but our arguments apply to other evidentiary statistical software as well. In discussing these limitations, we motivate our proposed framework that we introduce in Section~\ref{sec:framework}. 

\subsection{Source Code Review by the Defense}
Several groups have called for defense counsel to assess the validity of evidentiary software using source code review~\cite{bellovin2021seeking, lacambra2018opening, upturnAmicusBrief2020, pasquale2015black, chessman2017source, imwinkelried2016computer}. While examining the source code may help uncover programming mistakes, or ``miscodes,'' looking at the source code without also running the software executable may reveal little about the statistical tool's performance (e.g., accuracy or error rates), consequently leaving out a crucial metric for assessing the validity of statistical software output used in a case.\footnote{E.g.,~\citet{lander2016forensic} and~\citet{bellovin2021seeking} state that foundational validity requires that a method has been subjected to empirical testing that provides ``valid estimates of the method's accuracy''.} Illustrating this limitation in the context of PGS tools, scrutinizing the source code can verify whether the program correctly implements a fully continuous PG model. Without also having the ability to test the statistical software, however, defense counsel cannot fully determine the tool's performance (e.g., error rates) in any setting, including that of the defendant's case. Given this limitation, empirical testing is crucial for assessing the tool's performance and, consequently, for assessing the validity of the tool's output in any given case.

\subsection{Validation Studies by Relevant Scientific Communities}

Defense counsel may also use empirical assessment mechanisms to gauge the tool's performance. One such mechanism the criminal legal system already relies upon is validation studies. However, these studies face numerous limitations.

Many groups have rightfully pointed out that existing validation studies, even those that are peer-reviewed, do not rigorously assess evidentiary software because the authors or reviewers of such studies have financial or professional interests in ensuring that forensic laboratories use the software. For instance, amongst the eight TrueAllele validation studies that ~\citet{butler2021dna} enumerate, seven studies include Mark Perlin, the CEO of Cybergenetics Inc., which produces TrueAllele, as a co-author. The remaining study acknowledges Perlin and two Cybergenetics employees for their guidance and comments. In light of observations like these, many argue that independent groups with no stake in the outcome of validation studies should test the software~\cite{sinha2021radically, kwong2017algorithm, butler2021dna, upturnAmicusBrief2020, canellas2021defending, robinson_2021, lander2016forensic, kaye2011new, mnookin2010need}.

Even if more independent scientific groups were to validate evidentiary software, before conducting each study, the study designer must choose a set of test cases on which to evaluate the tool. As several commentators have pointed out, these chosen test cases are not guaranteed to align with the characteristics of any actual criminal case in which the tool is applied~\cite{lander2016forensic, butler2021dna, kroll165accountable}. Indeed, this is true regardless of whether the parties conducting the studies are independent of the developer or not. Consequently, the tool's performance and error rates reported in a validation study are not guaranteed to reflect the tool's validity when applied in the criminal legal system. Scholars have observed that evaluation studies of machine learning systems are often over-optimistic estimates of performance on new test cases; the performance of these systems can drop dramatically under slight variations of the test distribution~\cite{liao2021we, recht2019imagenet, cooper2021hyperparameter, buolamwini2018gender, zech2018variable, koenecke2020racial}.

In sum, validation studies require the designer to choose test cases and parameters that may not match those relevant to the case at hand. This mismatch can result in significant differences in the statistical software's performance and output, so testing the tool on inputs similar to the case at hand is imperative to ensure the tool, as applied in the specific case, is valid.

\subsection{Direct Testing by the Defense}
Instead of relying on other parties to empirically test the statistical software in contexts outside of any given case, defense counsel may directly test the tool by choosing inputs and running the executable program. Indeed, organizations like Upturn have joined defense attorneys in arguing in court for defense access to executable source code and additional resources to make this direct assessment~\cite{upturnAmicusBrief2020}, and many scholars have made similar arguments~\cite{lacambra2018opening, bellovin2021seeking, justiceInForensicAlgorithmicsAct}. We agree that defense access to the executable source code is crucial for assessing software validity. We argue, however, that current discussions of testing techniques are incomplete.
 
In particular, current discussions do not provide defense counsel with sufficient guidance to conduct such tests. While some groups have specifically suggested that the defense test software on inputs relevant to the defendant's case and have provided examples~\cite{bellovin2021seeking, lacambra2018opening}, these discussions do not provide any overarching framework that would equip defense counsel to make their determinations when navigating steps such as choosing inputs that are relevant to their client's case. To ensure that defense counsel can effectively test evidentiary statistical software, such guidance is necessary. \\
 
 \par \noindent Acknowledging the limitations of existing proposals for testing and of the other mechanisms we described in the preceding subsections, we propose one potential framework for testing statistical software, which we present in the next section.

\section{Robust Adversarial Testing}
\label{sec:framework}
We now present our framework for adversarial scrutiny of evidentiary statistical software.
Before we introduce our mathematical definition, we first provide the underlying intuition in Section~\ref{sec:defn}. In Section~\ref{sec:operationalizing}, we discuss how defense counsel can operationalize our definition.

As the previous section motivates, we would like to answer the question: \emph{Is the tool in question accurate on inputs corresponding to the defendant’s case}?
Accurate evidence from statistical software tools will aid the truth-seeking process of the criminal legal system and minimize the number of wrongful convictions.
It is well-known, however, that standard statistical validation provides no simple answer to this question because statistical validation makes claims about \emph{distributions of instances}, and not individual instances. Even if a statistical validation study establishes that a tool performs accurately across a broad distribution of instances, such a finding does \emph{not} imply any rigorous guarantees for a previously unseen instance relevant to the defendant's case.
For example, a probabilistic genotyping tool may be highly accurate on DNA mixtures containing genetic material from two individuals, but may fail to produce correct answers when the mixture has more or an unknown number of contributors. Similarly, gunshot detection tools may be highly accurate in otherwise quiet environments but have high error rates when the ambient environment is filled with noise from traffic, fireworks, or construction.

The difficulties surrounding the accuracy of statistical software can have direct consequences on the validity of evidence in court. Consider a hypothetical probabilistic genotyping tool that achieves perfect accuracy for 99\% of DNA mixtures, but is always wrong for the remaining 1\% of instances. If the defendant in question belongs to this 1\% of instances, the tool will be guaranteed to fail in this case despite its seemingly impressive accuracy.\footnote{Here and in other examples we use accuracy as a performance metric for ease of exposition. In an actual application of our framework, other performance metrics such as precision, recall, ROC curves, and others may be more meaningful depending on the tool and case in question.}

To formalize this intuition and ground the legal concept of adversarial scrutiny in the technical reality of statistical software, we build on conceptual advances from the literature on distributional robustness in machine learning and algorithmic fairness~\cite{hebert-johnson18a, dwork2021outcome,wald,biggio,robustopt,quinonero2009dataset,cooper2022accountability}. At its core, robust adversarial testing requires a tool to perform well on inputs most relevant to the defendant’s case. For instance, the accuracy of a DNA analysis tool on mixtures with two participants may have no relevance for a case where the mixture comes from five contributors. Determining precisely what input is relevant to a defendant’s case is a key step to operationalize our definition of adversarial scrutiny. Our definition empowers the defense to adversarially select inputs for testing the evidentiary software in question.

Next, we provide a definition of robust adversarial testing and then discuss how to operationalize this definition in practice. Intuitively, our definition decomposes assessing the performance of evidentiary software into two parts:
(i) the metric used to check performance, and
(ii) the distribution on which the metric is applied.
By allowing the defense to choose each component adversarially, our framework enables the defense to question evidentiary software in a way most relevant to the case at hand.

\subsection{Formal Definition of Robust Adversarial Testing}
\label{sec:defn}

We now state our model for adversarial testing. To ground this definition, it is helpful to keep in mind the example of probabilistic genotyping tools introduced in Section~\ref{sec:prob-genotyping}. A statistical tool computes a function $A\colon Z\to O,$ where $O$ is a set of outputs and $Z$ is a data universe of possible inputs. For a PGS tool, the output space $O$ may be the space of likelihood ratios, and the input space $Z$ is the space of DNA samples taken from the crime scene and the defendant.

For a given tool $A$ and distribution $Q$ over the data universe $Z$, let $\mathrm{check}(A, Q)\in\{\mathrm{Pass},\mathrm{Fail}\}$ be a test that determines whether the tool $A$ has \emph{acceptable} performance on the distribution $Q$. For instance, the distribution $Q$ may be a distribution over DNA mixtures with two contributors, and the check function may compute the error rate of a PGS tool $A$ on the distribution $Q$, returning \emph{Pass} only if the error rate is below some threshold. In a concrete validation study for a PGS tool, a finite dataset of test examples instantiates the distribution $Q$. The notion of “acceptable performance” references the idea of “general acceptance” in the law from Section 2. We envision that this check function can also be chosen adversarially by the defense to test for flaws that generally accepted evaluation metrics may hide. We will return to this point when we discuss how to operationalize the definition.

Further, let $F$ be a family of distributions over the same data universe $Z$. Concretely, this could be a collection of distributions over DNA mixtures with the same number of contributors as the sample in question, but perturbed values of laboratory parameters or quantities of DNA collected. Specifically, a family of distributions $F = \{Q_1, Q_2, Q_3\}$ could consist of three individual test distributions where: 
\begin{itemize}
    \item $Q_1$ is a distribution over DNA mixtures with two contributors and 100--200 pg of DNA.
    \item $Q_2$ is a distribution over DNA mixtures with two contributors where the DNA mixture was analyzed with lab equipment from a specific manufacturer.
    \item $Q_3$ is a distribution over DNA mixtures with two contributors and a contributor ratio between 2:1 and 4:1.
\end{itemize}
If the trial involves a DNA sample with the above characteristics (two contributors, 100--200 pg of DNA, lab equipment from the same manufacturer as in $Q_2$, and a likely contributor ratio between 2:1 and 4:1), each of the three distributions in the family $F$ stress tests the PGS tool in question on inputs that are relevant to the case. In addition, each distribution stress tests the PGS tool differently.
Hence the PGS tool's accuracy on these distributions is likely more representative of its accuracy on the defendant's inputs than the result of a validation study on a broad distribution over DNA mixtures that may only be superficially related to the DNA sample in question. As before, a dataset would instantiate each of the individual distributions $Q_i$ in a concrete adversarial evaluation during an individual criminal case.

With our formal setup in place, we now state our main definition:
\begin{definition}[Robust Adversarial Testing]
We say that a tool $A$ passes $(F, \mathrm{check})$-robust adversarial testing if for every distribution $Q$ in the family $F$ we have $\mathrm{check}(A, Q)=\mathrm{Pass}$.
\end{definition}

Importantly, since the performance check must pass for all distributions $Q$ in the family $F$, it must also pass for the worst-case distribution in the family. Therefore, the defense choosing the family $F$ corresponds to implicitly picking the worst-case distribution in the family on which to evaluate the tool. Our definition gives the defense broad latitude to choose the family $F$. However, nuance is required to select a family $F$ that is relevant to the defendant’s instance and avoids trivial examples, e.g., a point-mass on a known failure case.\footnote{Note that, if the defense were to choose \emph{irrelevant} inputs, the prosecution could likely have the results excluded from the case because "irrelevant evidence is not admissible." \emph{See, e.g.,} Fed. R. Evid. 402. Hence, the legal concept of relevance binds the defense selections in our framework. "Evidence is relevant if: (a) it has any tendency to make a fact more or less probable than it would be without the evidence; and (b) the fact is of consequence in determining the action." Fed. R. Evid. 401.} Selecting an appropriately rich and relevant family $F$ is one of the key challenges involved with operationalizing this definition.

Our definition makes use of a binary check function that ultimately outputs that a tool either ``passes'' or ``fails.'' One could also consider a continuous version of our definition where we simply report the worst-case metric over the distribution family. We elected to use a binary check function to abstract away questions of statistical validity (e.g., confidence intervals can be implicitly included in the check function) because judges and juries must ultimately make a validity determination for a particular tool in a particular case. For a binary check function, the choice of the threshold which constitutes ``passing'' is often sensitive and potentially contentious \cite{delgado2022uncommon}. In our definition, we empower the defense to select the threshold, and we envision that relevant research communities can empower the defense with guidelines for selecting this threshold. We discuss this in more detail below.

\subsection{Operationalizing Robust Adversarial Testing}
\label{sec:operationalizing}
We now discuss how to implement and operationalize our formal definition. We operationalize our definition in three steps. Because our definition is adversarial, we envision each step will be carried out by the defense and specific to the case at hand. The first step involves determining the data universe. The second step is the adversarial choice of a relevant distribution of inputs. Finally, the third step is the choice of a reasonable performance check to measure how well the tool performs on the chosen distribution of inputs.

\subsubsection{Determining the Data Universe}
The first step of our procedure is to determine what, precisely, is that data universe. Conceptually, one can think of the data universe as a large database enumerating all possible inputs to the system under study, background factors, system output, and ground truth. In the case of PGS tools, this ideal database may correspond to all possible DNA samples (with differing numbers of contributors, amount of genetic material, etc.), parameters of the analysis tool, and features of the defendant (ethnicity, age, etc.), and other such considerations. 

Such idealized databases are unavailable in practice. As a practical compromise, we envision the data universe as being the union of existing data from validation studies. For example, several hundred two, three, and four-person mixture samples are publicly available through the ProvedIt database from~\citet{alfonse2018large}, and ~\citet{butler2021dna} detail more than 60 validation studies of varying size testing PGS tools under different input conditions. The extent to which existing databases are complete and accessible is an ongoing challenge. However, the lack of representative data in existing databases may itself indicate that the tool has not been sufficiently validated.

\subsubsection{Adversarial Selection of the Input Distribution}
After specifying the data universe, the next step of our adversarial procedure requires the defense to choose the family of distributions on which we evaluate the tool. The choice of distributions is not just specific to the problem the software maker purports to solve. It is also specific to the defendant. The defense will choose the distributions to reflect salient information available about the defendant's case.

For example, if the DNA mixture in the defendant’s case likely involves more than three contributors and trace amounts of DNA, then the defense might evaluate the performance of the tool on other mixtures with more than three contributors and similar quantities of DNA.
As a second example, consider a hypothetical software tool for analyzing videos, e.g., to determine whether a specific individual appears in a given video segment. If the tool analyzed video footage of a supposed crime that occurred at night, where the accuracy of computer vision algorithms may be worse, it is reasonable for the defense to interrogate the performance of the tool specifically on nighttime video. The method of determining the input distributions necessarily depends on context.

We propose that the tool should have acceptable performance at least on instances sharing some number of features in common with the defendant’s instance. For PGS tools,~\citet{butler2021dna} describe more than 20 distinct features that contribute to the performance and validity of these systems: the number of mixture components, the quantity of DNA, and so on. In this case, we would evaluate the tool on distributions of instances that match some number of the defendant's features (e.g., the same number of mixture components and the same quantity of DNA). Formally, this corresponds to taking the distribution family to be all $k$-way marginals defined by the defendant's features.

To implement this procedure, we start with the database of all instances described in the previous step. The defense then defines each possible distribution by first picking one or more features, for instance, the amount of DNA in the sample, that are deemed relevant.
Then, we consider the defendant's value or range of possible values for this feature, say 100 picograms, and take all instances in the database that approximately match this feature and have between, say, 95-105 picograms of DNA. This defines one particular distribution. The family of distributions then arises from repeating this process multiple times and also considering feature interactions (e.g., matching on both DNA quantity and the number of mixture contributors).

Determining which features are relevant, the number of features to consider jointly, and the number of distributions to include in the family ultimately depends on the context involving both the tool in question and the case at hand. These choices present a trade-off between requiring more specific validation of the evidentiary software and placing an unreasonably large burden on its validation that would preclude the use of all statistical software (e.g., if the family of distributions is too large or involves too specific distributions).

\subsubsection{Performance Check Selection}

The choice of a performance check relies on the defense choosing an evaluation metric that is appropriate for the particular tool and instance at hand. For instance, \citet{lander2016forensic} recommends using the inclusion error rate to evaluate a variety of forensic procedures. In general, the choice of a performance check should incorporate the notion of "general acceptance" borrowed from the legal standard, with specific attention to the possibility that a method will fall out of general acceptance. Concretely, the relevant legal and scientific research communities can keep track of what is considered acceptable for a task at hand. This empowers relevant research communities to remove old standards from general acceptance as acceptable performance evolves. This could happen, for example, by publishing results with improved accuracy or better trade-offs between false positives and negatives, which the defense can then point to when establishing the standard used in the check function.  For instance, a software tool that recognizes handwritten digits with 80\% accuracy no longer has general acceptance, since the task has long been solved with significantly higher accuracy~\cite{lecun1998gradient}.

If a suitable performance measure is unavailable, it is a good indication that the task or problem formulation themselves have not gained general acceptance, or have fallen out of general acceptance. This, on its own, is reasonable cause for contestation of the output of the tools. Take for instance the hypothetical case of a prosecution team that wishes to introduce evidence based on a risk score that rates the “latent criminality” of a person based on facial images. As the task lacks a scientific basis, there is no performance measure for the task that has gained “general acceptance” by the relevant scientific communities.

On a more technical note, the performance check abstracts out issues of statistical significance. If the test distributions chosen by the defense correspond to a sub-population that is too small to estimate the performance with high confidence, then the check fails.

\subsection{Relationship to Existing Lines of Work}
Our proposal builds upon and benefits from recent lines of work in several different communities across law, forensic science, and machine learning.

\vspace{2mm}
\noindent \textit{Statistical Evidence.}
There is a long line of work in the legal community discussing the use of statistical evidence in court \citep{referencemanual,scientificevidence}. Closely related to our paper, \citet{g2i} discuss difficulties that arise when courts apply statistical statements to individuals. The authors call this step ``group to individual'' (G2i) inference and suggest standards of evidence for such inferences. At a high level, our concerns with statistical evidence are similar, but our perspective on the role of software differs markedly: \citet{g2i} consider risk scores from a risk assessment tool as sufficient to establish error rates for evidence applied to individuals. In contrast, we question the validity of such risk scores and propose adversarial testing of statistical software as a way to test the risk assessment tool in question.

Another line of work in the statistical evidence literature considers the subtleties involved in choosing an appropriate ``reference class'' when making probability estimates about an individual case. Technically, this corresponds to choosing which variables to condition when fitting a statistical model. \citet{allen2007problematic} argue the difficulties involved in choosing an appropriate reference class limit the value of statistical models in court. \citet{cheng2009practical} suggests using ideas from model selection to address the reference class selection problem. While choosing a reference class bears some similarities to choosing our distribution family, these problems differ in an important aspect: The reference class problem is about how to build probability models, e.g., by appropriately selecting the features in a regression, whereas we are concerned with using the distribution family to validate already-fit models in software.

\vspace{2mm}
\noindent \textit{Robust Machine Learning.}
Our proposal benefits from close ties to recent lines of work in machine learning.
The first is a close connection to a robustness notion that originated in work on algorithmic fairness, but has since found numerous other applications. \citet{hebert-johnson18a} argue that existing fairness criteria are too weak insofar as they only hold in the population as a whole but not in rich families of sub-populations. This observation motivated the notion of multi-calibration and multi-accuracy that require a statistical model to achieve calibration and accuracy, respectively, in a given family of sub-populations. Subsequent work by~\citet{dwork2021outcome} introduced a closely related notion called \emph{outcome indistinguishability} with an emphasis on the perspective of reasoning about statistical models on individual instances. It is not difficult to show that our notion of robust adversarial testing generalizes multi-calibration and multi-accuracy at a formal level. But it also shares normative content. All notions aim to strengthen broad average measures of performance with stronger guarantees that come closer to providing meaningful guarantees for individual instances.

Fundamentally, robust adversarial testing requires a tool to perform well on relevant distributions related to the reference population. This requirement accounts for a central failure point of statistical machinery. An increasingly rich literature in the machine learning community has documented the failure of statistical models to generalize from one domain to another closely related domain \citep{biggio,quinonero2009dataset,torralba2011unbiased,zech2018variable,recht2019imagenet,wilds}.

\subsection{Discussion and Limitations}

Our definition establishes an important bridge between legal practice and machine learning research on distributional robustness. This connection provides a beneficial connection in both directions. On the one hand, criminal defense advocates can more effectively challenge tools using insights from lines of investigation that exhibit failure points of existing machine learning techniques. On the other hand, as machine learning experts work to make tools more robust to changes in distributional contexts, we expect that more well-designed evidentiary statistical software will hold up against defense challenges by our standard.

There remains a significant burden on the defense in at least two regards. First, our framework requires the tool to solve a well-defined statistical problem. If the problem formulation underlying the software tool is flawed, this on its own should become the primary point of contestation. Defense counsel must therefore be able to identify software that fails with regard to problem formulation. Difficulty in choosing a well-defined data universe, distribution family, or performance measure is a helpful indicator that a problem formulation may be invalid. These are, however, imperfect indicators that may miss other challenges in the problem formulation \cite{abebe2020roles,passi2019problem}. For instance, scholars argue that risk assessment tools conflate different notions of risk. Namely, they can conflate missing a court appointment with absconding from the jurisdiction or violently harming another person before a trial can be held \cite{abebe2020roles,koepke2018danger,passi2019problem}.
Moreover, in some cases the problem statement may lack a scientific basis entirely \cite{sloane2021algorithmic,sloane2022silicon}.

Second, our proposal leaves open aspects of executing our definition in context. There may be disagreement on what inputs and performance measures are relevant. The defense may not have access to relevant test cases or relevant data may be unavailable. The lack of representative data in existing databases may itself indicate the tool has not been sufficiently validated for the case at hand. The right level of adversarialism is also important. A defense that is too adversarial might choose a set of guaranteed failure cases, in which case the prosecution would likely argue that this choice was unreasonable. Our framework does not resolve what happens when the prosecution contests the choices of the defense.

While these are important challenges, attempts to apply our framework would provide valuable transparency about where concretely the process of adversarial scrutiny ran into difficulty. These challenges also point to fruitful research directions. Additional research can help in identifying misapplications of statistics, as well as facilitate the choice of parameters in our definition.

\section{Confronting Structural and Institutional Challenges}
\label{sec:barriers}

The use of evidentiary software without adequate scrutiny distorts the truth-seeking process and undermines fairness in the U.S. criminal legal system. In this work, we propose robust adversarial testing as a technical audit framework to empower criminal defense counsel to adversarially scrutinize evidence output by statistical software. This framework provides an approach to mitigate inadequate validity testing and cross-examination of forensic evidence. In doing so, we take a step towards formalizing an important aspect of audits of evidentiary statistical software.

Our framework can apply broadly throughout the life-cycle of a case, from plea negotiations to admissibility hearings to trial and beyond. It can be useful to the defense regardless of the varying legal standards that apply at each stage of the case, such as the more lenient threshold for admissibility or the more onerous standard of proof at trial. Further, while we focus on statistical software used as evidence, our approach could also apply to statistical software used for other criminal legal purposes. For instance, our framework could be used to scrutinize face matching software that police rely on during investigations, or to scrutinize risk assessment instruments that decision makers use to predict recidivism during pretrial bail hearings, sentencing, and parole. 

While our work is focused on statistical software, we also believe that such frameworks could be provided for non-statistical software, such as hash-matching algorithms used to identify contraband files or malware tools used to conduct remote investigations of networked digital devices. We encourage academic research and relevant technical communities to investigate and formulate similar adversarial frameworks in such cases. 

Notwithstanding the potential for this framework to have broad applicability, a host of structural and institutional barriers challenge the implementation of our framework in practice. Without simultaneously tackling these barriers, no framework can address the real needs of defense counsel or lead to long-lasting changes. Below, we highlight some of these challenges that currently create a barrier for proper evaluation of evidentiary software. We focus on two key axes relevant to our framework: (1) defense lack of access to software and data, and (2) defense lack of access to expert witnesses. We then discuss proposals for mitigating these barriers and highlight avenues for further consideration.

\vspace{2mm}
\noindent \textit{Access to Software and Data.} Vendors of statistical software can severely limit access to the software, for instance, by citing concerns around intellectual property or by using contract law to bar independent audits. Specifically, multiple vendors have asserted trade secret law to block criminal defense subpoenas and discovery requests for source code and other information about how their software works~\cite{katyal2021trade, siems2022trade, wexler2018life, ram2017innovating}. Meanwhile, some vendors also refuse to grant research licenses to independent academic scientists who wish to evaluate their tools, even while claiming to judges across the country that their tools are subject to peer review and hence should be admissible as evidence in court~\cite{howard_2021}. This practice runs directly contrary to scientific values of peer review and reproducibility, and presents one explanation for the dearth of independent peer-reviewed validation studies of evidentiary statistical software~\cite{lander2016forensic}.

Even when vendors do offer defense counsel some access to aspects of their software, they may demand onerous terms and conditions that make the access largely ineffective for performing meaningful assessments. For instance, in one case involving a popular probabilistic genotyping software, TrueAllele~\cite{truealleleProduct}, the vendor initially demanded that the defense team only access the source code using a read-only iPad and only use pen and paper to take notes. Only after extensive lawyering by a coalition of skilled attorneys from across the country---a luxury that most criminal defendants lack---was the defense team able to obtain more reasonable terms of access~\cite{kirchner_2021, 2021state}.  

We echo the prior arguments in \citet{wexler2018life} that, at minimum, judges should not extend trade secret law to block criminal defense access to relevant information about how evidentiary software works. Regardless of the precise information the defense is seeking to obtain---source code, validation studies, executable versions of software programs, etc.---vendors' intellectual property interests can be protected to the full extent reasonable by ordering disclosure subject to appropriate protective orders. Trade secret law offers no good reason to withhold such information from the defense.\footnote{Lawmakers have begun taking important steps to limit vendors' ability to assert trade secrets to block criminal defense scrutiny. See, for instance, the Justice in Forensic Algorithms Act of 2021, H. R. 2438 (2021).}

Broadly, vendor-imposed limits on defense access to evidentiary software reflect more widespread barriers to criminal defense access to data. For example, national DNA databases are available to law enforcement to search for evidence of guilt, but are often unavailable to defense counsel to search for evidence of innocence~\cite{garrett2021autopsy_pg189}.  The same is true for fingerprint~\cite{garrett2021autopsy_pg192} and face matching databases~\cite{mak_2019}. Judges have interpreted federal law to make private sector digital communications content databases available to law enforcement but categorically barred to the defense~\cite{wexler2020privacy}. We argue that these access limitations should be reversed; There are no justifiable reasons to categorically deny defense counsel access to relevant data that they need to investigate forensic evidence in their case.  

Without access to data to run evidentiary statistical software tools, defendants are at a disadvantage in both testing the software to expose flaws in the government’s evidence of guilt, and in using the software to surface evidence of innocence. For adversarial scrutiny to succeed, defendants need access to relevant evidence, whether that means queries of forensic databases, software source code, training data, or executables to run and test the software.

\vspace{2mm}
\noindent \textit{Access to Technical Experts.} Even if defense counsel manage to obtain meaningful access to software and data, they face another substantial barrier to performing adversarial scrutiny: access to experts. Defense counsel themselves are typically not trained to do the kind of technical work required to scrutinize statistical evidentiary software, and most defense offices lack the resources to support specialized attorneys or staff with these skills. Hence, most defense counsel will need to depend on outside experts. Yet, while the Supreme Court has held, at minimum, that indigent defendants in capital cases have a due process right to state-funded psychiatric experts,~\cite{1985ake} case law remains ambiguous as to whether indigent defendants have a right to expert assistance in other contexts~\cite{giannelli2003ake}.

Further, the structure of the forensic science software market tilts the pool of potential experts in favor of the prosecution, since law enforcement and prosecution entities are the primary paying customers---if not the developers themselves---and private developers build tools primarily to serve law enforcement needs~\cite{ram2017innovating, siems2022trade}. Specifically, employees of software vendors are ideally positioned to gain expertise as to their own products, and sometimes testify about the strengths of their tools on behalf of the prosecution.\footnote{\emph{See, e.g.}, https://www.cybgen.com/information/newsroom/2016/apr/files/B-Perlin-second-declaration.pdf.} However, these employees can hardly be expected to volunteer information about the weaknesses of their tools for use by the defense. To fill this gap, defense counsel may seek to rely on academic experts. But here again they face obstacles. Defense counsel have limited opportunities to identify the right academics and to provide them with sufficient incentives and training. In recent years, there have been efforts, such as the nonprofit, PDQuery~\cite{pdquery_2021}, which seek to address this problem by providing a matching service to connect technical expert volunteers with public defenders. More innovations of this sort are needed at scale.

In sum, defense counsel now confront a growing number and diversity of evidentiary statistical software used in an increasing number of cases without access to the technical expertise they need to adequately evaluate these tools.

\subsection*{A Way Forward}

Beyond the reforms mentioned above---specifically, eliminating trade secret barriers for criminal defense access to relevant evidence, ensuring defense access to relevant data, and scaling existing efforts to match criminal defense counsel with volunteer academic experts---we offer three additional proposals to help mitigate the structural and institutional barriers to implementing our framework for adversarial scrutiny of evidentiary statistical software.

First, our adversarial scrutiny framework provides an opportunity for the computational academic community to contribute in ways beyond serving as experts in individual cases. Namely, academic scientists could create, sign, and publicly distribute generic affidavits that are ready to file in court and that attest to known issues from existing research and the current state-of-the-art for various evidentiary statistical software tools. The affidavits could support the selection of the input distribution and performance check in our framework using existing research and known limitations of the task that the software purports to perform. The affidavits could also do the challenging work of translating expert knowledge to make it accessible to generalist judges and lay jurors. In short, these affidavits could provide an off-the-shelf resource for defense counsel who are scrutinizing the software in any specific case.

Second, academic scientists could create off-the-shelf testing tools that would allow for checking for different cases in a more streamlined way, requiring less expertise and time from the defense. Rather than starting from scratch for each case, defense counsel could instead use an existing tool corresponding to a particular evidentiary software to enter the most relevant inputs to their case at hand and test for the tools' performance. An important component of such testing tools enabling adversarial scrutiny would also be comprehensive databases of inputs (e.g., DNA samples) so that the defense can easily select relevant subsets of the data to stress test the statistical software in question.

Importantly, our proposal for scientific communities to provide off-the-shelf affidavits and testing tools does not serve as a substitute for defense offices having the appropriate expertise and time to conduct adequate scrutiny, but aims to ease the burden on these offices by sharing information that is currently present in scientific communities.

Third, we advocate for government funding for public defender offices to hire specialized technical staff who can perform robust adversarial testing in-house. The Digital Forensics Unit (DFU) of The Legal Aid Society of New York City offers a ready model. The DFU integrates technologists and defense attorneys in a single unit to enable coordinated advocacy across computer science and law~\cite{legalaid}.

Prior commentators have emphasized the importance of independent audit and oversight bodies to help ensure the fair and accurate use of new AI and other software products in high stakes settings such as the criminal legal system. We agree that such independent audit and oversight bodies are essential to improve the quality of forensic evidence in the courts. This includes government oversight bodies that take a neutral role, such as the National Institute of Standards and Technology and the recently disbanded National Commission on Forensic Science~\cite{mnookin2018uncertain}.

Yet, oversight committees tasked with \emph{neutral} evaluations of forensic science methods are not ideally situated to perform the kinds of stress-tests we envision for \emph{adversarial} scrutiny of evidentiary statistical software. Rather, public defender offices provide a natural home for technical staff to perform adversarial scrutiny because these offices are already committed to zealous advocacy on behalf of the accused. We hope that placing technical auditors inside public defender offices will protect against industry capture. Further, by placing technical staff inside public defender offices, we directly empower defense lawyers and their clients, who often belong to the most marginalized communities that are disproportionately harmed by faulty forensics~\cite{sinha2021radically}.

\begin{acks}
JM was supported by the National Science Foundation Graduate Research Fellowship Program under Grant No. DGE 1752814. The authors declare no other financial interests. 

The authors would like to thank the following individuals for invaluable discussions and comments on previous drafts of this work: 
Pieter Abbeel, 
Nathan Adams, 
Steven M. Bellovin, 
Matthew Boulos, 
Alex Carney, 
Edward Cheng, 
A. Feder Cooper, 
Fernando Delgado, 
Brandon L. Garrett, 
Jonah Gelbach, 
Andreas Alexander Haupt,
Jon Kleinberg, 
Logan Koepke,
Beryl Lipton,
Deborah Raji,
Andrea Roth, 
Christopher Strong,
Ramon Vilarino,
Dana Yeo,
Harlan Yu, as well as members of the Berkeley Algorithmic Fairness and Opacity Group (AFOG) and the Berkeley Equity and Access, in Algorithms, Mechanisms, and Optimization research group (BEAAMO). The authors would also like to thank the anonymous reviewers for their feedback. This work would not have been possible without insightful discussions with Richard Torres and Clinton Hughes from the Brooklyn Defenders and Queens Defenders offices, Elizabeth Vasquez from Legal Aid, as well as Nathan Adams and Dan Krane from Forensic Bioinformatics. The authors also received substantial reference support from Dean Rowan at the Berkeley Law Library. 
\end{acks}

\bibliographystyle{ACM-Reference-Format}
\bibliography{sample-base}

\clearpage
\newpage

\begin{table*}[!h]

  \begin{tabularx}{\linewidth}{lX}
    \toprule
    Category&Products\\
    \hline \hline
    Breath Testing & Alcotest 9510~\cite{alcotest9510}, Intoxilyzer 8000~\cite{intoxilyzer8000}, LX9 Breathalyzer~\cite{lx9breathalyzer}\tabularnewline \hline
    Digital Device Extraction Forensics & Cellebrite Universal Forensic Extraction Device (UFED)~\cite{cellebriteUFED}, GrayKey~\cite{graykey}, Magnet AXIOM~\cite{magnetAXIOM}\tabularnewline \hline
    Environment Audio Detection & SENTRI (Sensor Enabled Neural Threat Recognition and Identification)~\cite{sentri}, ShotSpotter~\cite{shotspotter}\tabularnewline \hline
    Probabilistic Genotyping & FST (Forensic Statistical Tool)~\cite{fstUse}, STRMix~\cite{strmixProduct}, TrueAllele~\cite{truealleleProduct}\tabularnewline \hline
    Toolmark Analysis & Correlation Engine~\cite{correlationEngine}, Matchpoint~\cite{matchpoint}\\
  \bottomrule
\end{tabularx}
\caption{Example categories of evidentiary software and a sample of known products for each category.}
\label{tab:software}
\end{table*}

\appendix

\section{Example Evidentiary Software}
In Table~\ref{tab:software}, we present examples of different categories of evidentiary software as well as examples of known products for each such category.

\end{document}